# Tunable Graphene Split-Ring Resonators

Qiaoxia Xing[1,2], Chong Wang[1,2], Shenyang Huang[1,2], Tong Liu[1,2], Yuangang Xie[1,2], Chaoyu Song[1,2], Fanjie Wang[1,2], Xuesong Li[3,4]*, Lei Zhou[1,2]* & Hugen Yan[1,2]*

[1]State Key Laboratory of Surface Physics, Department of Physics, Fudan University, Shanghai 200433, China.

[2]Key Laboratory of Micro and Nano-Photonic Structures (Ministry of Education), Fudan University, Shanghai 200433, China

[3]State Key Laboratory of Electronic Thin Films and Integrated Devices, University of Electronic Science and Technology of China, Chengdu 610054, China

[4]School of Electronic Science and Engineering, University of Electronic Science and Engineering Technology of China, Chengdu 610054, China

*Corresponding authors: lxs@uestc.edu.cn (X. L.), phzhou@fudan.edu.cn (L. Z.), hgyan@fudan.edu.cn (H. Y.)

**A split-ring resonator is a prototype of meta-atom in metamaterials. Though noble metal-based split-ring resonators have been extensively studied, up to date, there is no experimental demonstration of split-ring resonators made from graphene, an emerging intriguing plasmonic material. Here, we experimentally demonstrate graphene split-ring resonators with deep subwavelength (about one hundredth of the excitation wavelength) magnetic dipole response in the terahertz regime. Meanwhile, the quadrupole and electric dipole are observed,**

**depending on the incident light polarization. All modes can be tuned via chemical doping or stacking multiple graphene layers. The strong interaction with surface polar phonons of the $SiO_2$ substrate also significantly modifies the response. Finite-element frequency domain simulations nicely reproduce experimental results. Our study moves one stride forward toward the multi-functional graphene metamaterials, beyond simple graphene ribbon or disk arrays with electrical dipole resonances only.**

## I. INTRODUCTION

Metamaterials have been explored extensively for decades owing to the potential applications including negative refraction and superlens which need both electric and magnetic responses at optical frequencies [1-3]. The split ring resonator (SRR) is a key element of metamaterials to produce circular induced current, leading to a resonant magnetic response at optical frequencies [4-7], which is unattainable in natural materials [5]. Though widely studied, the performances of conventional noble-metal SRR are hampered by the high loss, poor electromagnetic wave confinement and the lack of efficient ways to tune the spectral response [8-9]. Recently, graphene, with intriguing properties including high mobility, strong light-matter interaction and excellent chemical stability, has been considered as an alternative material to design SRRs [10-13]. Graphene has two unique properties for the application in SRR: (i) the carrier density of graphene can be electrically, chemically or optically tuned [14-17], and (ii) the relatively low carrier density of graphene leads to stronger mode confinement when compared with noble metals [8, 18]. Such merits already make graphene an ideal platform to host tunable terahertz (THz) plasmons [8, 14, 19-20]. Even though the fundamental nature aforementioned and several simulation studies have verified the feasibility of using graphene SRR to exhibit tunable magnetic response at THz frequency [10-11, 21], the experimental study remains elusive. Only recently, complementary graphene SRR (SRR voids) arrays were studied [22].

In this work, split ring resonators (SRRs) based on chemical vapor deposition (CVD) graphene are fabricated by conventional electron beam lithography (EBL) [23]. Tunable resonances, including the magnetic dipole resonance (a term consistent with the literature) [7, 24], are systematically studied. The tunability by chemical doping, graphene layer number control and coupling to surface polar phonons is demonstrated. Such versatility is further confirmed by simulations, which perfectly match the experimental results.

## II. EXPERIMENT

### A. Sample preparation and characterization

SRR arrays with area of 500 um × 500 um were fabricated on double-side polished float-zone intrinsic silicon with or without 285 nm native oxide layer. CVD grown graphene was transferred with PMMA (Poly(methyl methacrylate)) support to the aimed substrates after wet-etching of the copper film by $FeCl_3$ [25]. Double-layer (trilayer) graphene was obtained by two (three) successive transfers of single-layer graphene. It is worth noting that such transfer procedure could not lead to the modification of the band structures because of the random orientation of domains and relatively uncontrolled interlayer separation of the artificially stacked layers [26]. Graphene SRRs were patterned with 20kV electron beam lithography using AR-N 7520 (ALLRESIST) as a negative electron beam resist, and were etched in oxygen plasma after development. Chemical doping was performed by exposing the samples to nitric acid vapor for 15 min [16, 19]. Baking the doped sample at 100 ℃ with different time duration could reduce the doping to various concentrations. The chemical doping is stable in air at room temperature for an extended period. The transmission spectra of the arrays were acquired by a Fourier-transform infrared spectrometer coupled to an infrared microscope (Bruker Vertex 70V and Hyperion 2000), in conjunction with a broadband far-infrared polarizer. A liquid-helium-cooled silicon bolometer (IR Labs) served as the detector of far-infrared light. All the measurements were performed at room temperature with samples and the external beam path in the nitrogen environment to minimize the water vapor absorption. These CVD samples are relatively stable, as evidenced in FIG. S8 of the Supplemental Material [27].

### B. Numerical simulations

The transmission spectra were calculated with a frequency-domain solver of Maxwell equations based on the finite element software (Comsol Multiphysics). The periodic split ring structure was simulated using Floquet-Bloch periodic boundary conditions at *x*- and *y*-directions. Perfectly matched layers were imposed along the *z*-direction. The proper convergence of simulation was ensured by monitoring the transmission coefficient while the meshing was refined in an iterative process. Graphene was modeled as a two-dimensional layer with a complex AC conductivity described by the Drude mode:

$$\sigma_s = \frac{iD}{\pi(\omega + i\gamma)}$$

where $D$ is the Drude weight, $\gamma$ is the carrier scattering rate and $\omega$ is the frequency. Including the interband contribution is essential when $\hbar\omega > 2E_F$. In our case, however, interband conductivity was not taken into account due to the relatively low frequency of resonances. Previous results have confirmed that the resonance of multi-layer graphene patterns is equivalent to that of a single layer with a greater Drude weight [28]. Thus larger Drude weight was used in the simulation to represent more graphene layers in our study. For the $SiO_2$/Si substrate, the frequency-dispersive dielectric function of $SiO_2$ was used in the simulation. See the details of the simulation in the Supplemental Material [27].

## III. RESULTS AND DISCUSSION

Fig. 1(a) shows the far-infrared transmission measurement scheme. Linearly polarized light is either parallel ($E_{//}$) or perpendicular ($E_\perp$) to the symmetric axis of SRR. Extinction spectra 1-$T/T_0$ characterize the electromagnetic responses of the SRRs, where $T$ and $T_0$ are the transmission of the quasi-normal incident light through SRR arrays and the bare substrate, respectively. Fig. 1(b) shows a typical scanning electron microscopy (SEM) micrograph of a SRR array, with each SRR dimension of ~400 nm. Fig. 1(c) displays the Drude response of successive stacked CVD graphene from one to three layers before nanofabrication. These extinction spectra are well described by the Drude model, as shown by the fitting curves in Fig. 1(c). Stacking multiple layers effectively increases the Drude response amplitude and provides more pronounced resonances after nanofabrication of SRRs [19, 28]. Fig. 1(d) presents the extinction spectra of a typical SRR array on Si substrate. Salient resonant peaks can be observed: one peak $P_1$ for parallel polarization and at least two peaks $P_2$ and $P_3$ for perpendicular polarization.

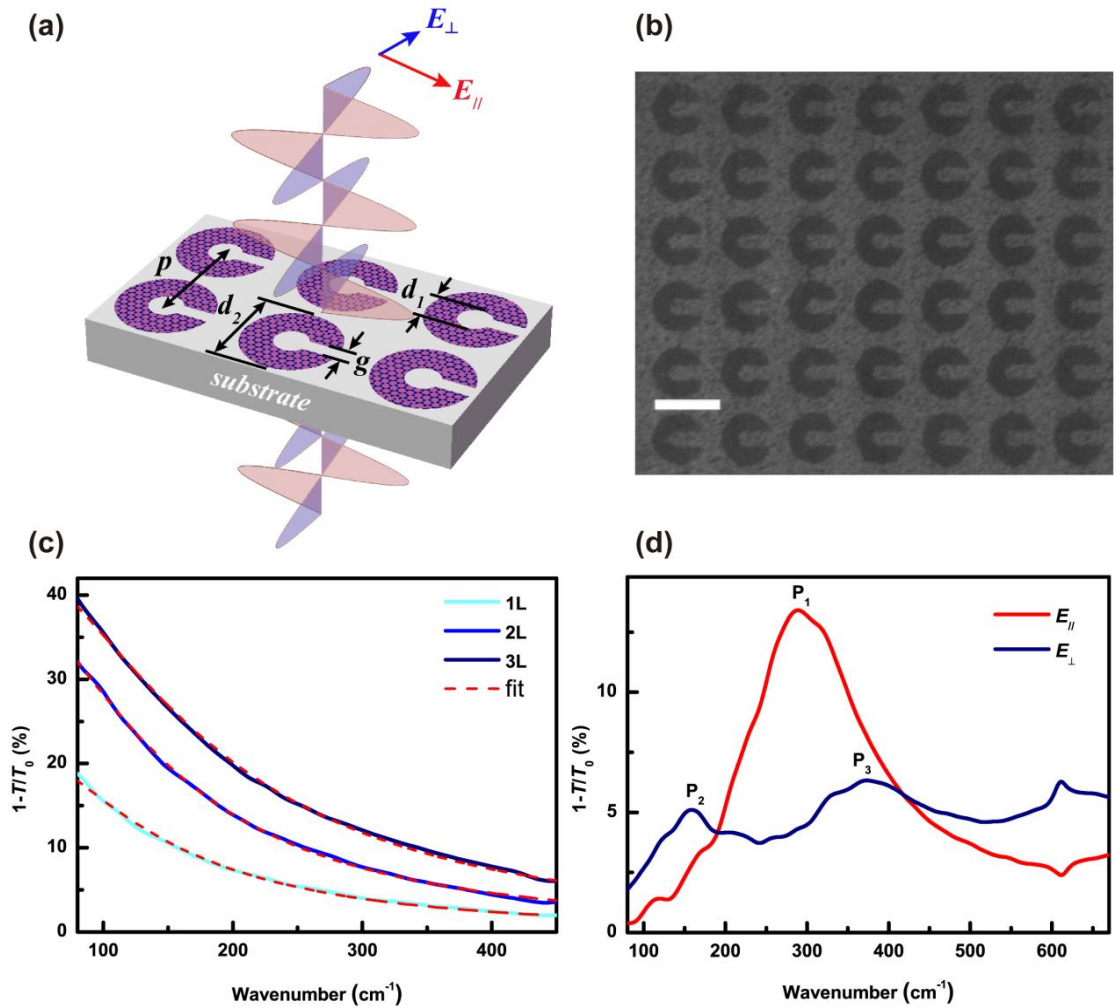

**FIG. 1. Characterization of graphene layers and SRRs. (a) Schematic of a typical SRR array with normal incident far infrared radiation. The incident light is polarized either parallel ($E_{//}$) or perpendicular ($E_\perp$) to the symmetry axis of the split ring. A square lattice array of the**

**SRRs made of CVD graphene is located in the xy plane. Geometric parameters are indicated: the ring outer diameter $d_2$, inner diameter $d_1$, a gap $g$ and the period of the lattice $p$. (b) An SEM micrograph of a SRR array with parameters of $d_1$ = 100 nm, $d_2$ = 400 nm, $g$ = 100 nm and $p$ = 500 nm. The scale bar is 500 nm. (c) Drude response of one to three-layer graphene on Si/SiO$_2$ substrate. Dashed lines are corresponding fitted curves using Drude model with fitting parameters of Drude weight $D$ and scattering rate $\gamma$. Their values ($D$ and $\gamma$) are $8.2 \times 10^{10} \, \Omega^{-1} s^{-1}$, $112 \, cm^{-1}$; $1.6 \times 10^{11} \, \Omega^{-1} s^{-1}$, $106 \, cm^{-1}$; $2.2 \times 10^{11} \, \Omega^{-1} s^{-1}$, $125 \, cm^{-1}$ for one to three layers respectively. (d) Typical extinction spectra of an SRR array on Si substrate with parameters the same as that in the SEM image in (b). Red and blue curves are for the parallel and the perpendicular polarizations, respectively.**

### A. Far infrared response of graphene SRRs

In our systematic study, we first characterized the response of the SRR arrays on highly resistive Si in the far-infrared region in detail. Here, all the patterned graphene samples are double-layer, since the optical response of single-layer graphene SRR is relatively week on such a high permittivity substrate. As presented in the left part of Fig. 2(a), prominent resonance peaks can be observed in both polarization configurations; two peaks can be resolved as the polarization is perpendicular to the symmetric axis of SRR (blue curves), while only one peak is observed with parallel polarization (red curves). By changing the outer diameter $d_2$ from 300 nm to 400 nm (other geometrical parameters $d_1$ = 100 nm, $g$ = 100 nm and the spacing between the unit cells $p$-$d_2$ = 100 nm are all fixed), all the resonance peaks redshift. A further redshift is shown by a sample with an outer diameter $d_2$ = 500 nm and an inner diameter $d_1$ = 160 nm. This is as expected since localized resonances typically soften with increasing structure size. Measured spectra agree excellently with the simulation results shown in the right part of Fig. 2(a). When the electric field is polarized along the symmetric axis of the SRR, only the dipole like resonance is preferentially excited due to the preserved ring symmetry. However, the similar symmetry no longer exists for perpendicular polarization, giving rise to other higher-order oscillation modes, among which the mode with higher frequency is attributed to the quadrupole (the mode component is not pure, but the assignment is consist with the literature) [24, 29-31], while the nature for the mode with lower frequency can be found clues through the experiment shown in Fig. 2(b). These spectra characterize the SRRs fabricated from the same piece of graphene and with the same geometrical parameters ($d_1$ = 170 nm, $d_2$ = 400 nm, $p$ = 500 nm) but different split gap $g$. As $g$ increases form 70 nm, 100 nm, to 150 nm, the frequencies of $P_2$ and $P_3$ modes in the perpendicular polarization slightly blueshift. However, the peak intensity of $P_2$ is apparently enhanced, which is in sharp contrast to the quadrupole ($P_3$) mode, as shown in Fig. 2(c). Thus we conclude that this lower frequency mode $P_2$ is very sensitive to the split gap and this mode should be the magnetic dipole mode which features a circular current. In SRRs, an open gap provides a capacitance $C$. The resonance frequency and amplitude of the magnetic mode increase as $C$ decreases, which explains our experimental findings. The experimental results can be reproduced by the simulation, as shown in the right part of Fig. 2(b), which captures every detail of the measured spectra.

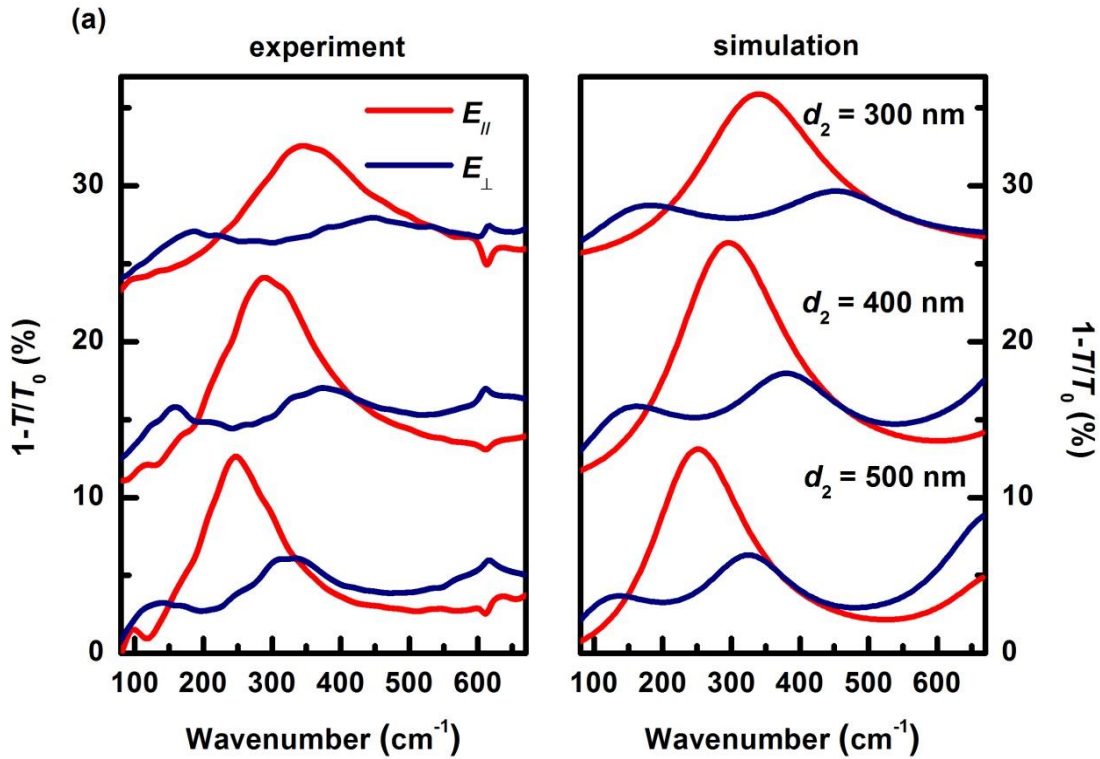

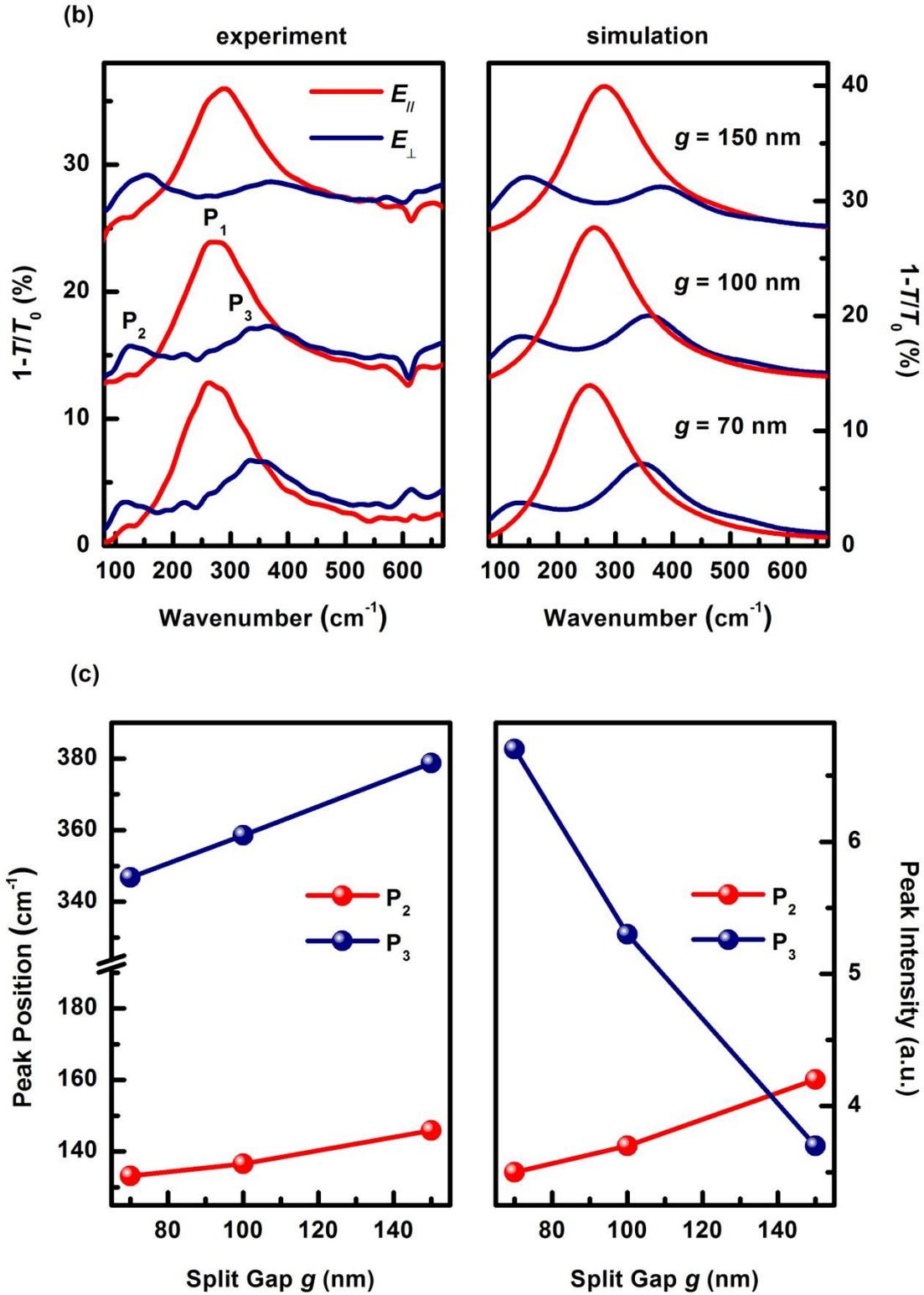

**FIG. 2.** SRRs on Si substrate with varied diameter and split gap. (a) Measured and simulated extinction spectra for the SRRs with varied outer diameter. Both parallel and perpendicular polarization spectra are shown. Spectra are vertically offset for clarity. (b) Measured and simulated extinction spectra for the SRRs with increasing split gap from 70

nm, 100 nm to 150 nm. Spectra are vertically offset for clarity. (c) Peaks positions and intensities of peak $P_2$ and peak $P_3$ as a function of the split gap *g*.

## B. Electromagnetic simulation results

To fully reveal the mode nature of the extinction peaks in Fig. 2, we further analyze a typical resonance and their near-field distributions with full-wave electromagnetic simulations using Comsol Multiphysics. See the details in Experimental section. Fig. 3(a) shows the simulated extinction spectra of the SRR array whose measured spectra are presented in Fig. 1(d). Corresponding surface electric field magnitude distribution (upper panels) and the *z*-direction magnetic field magnitude distribution as well as the surface current density (lower panels) are shown in Fig. 3(b). For the resonance peak $P_1$, the surface current density distribution, which is symmetric with respect to the symmetry axis of SRR, reveals the nature of an electric dipole mode since the electric field polarization is parallel to the symmetric axis. When the electric field polarization is perpendicular to the axis, an electric dipole forms along the external field direction, as represented by the electric field distribution of peak $P_2$. Therefore, the induced circular surface current flows along the whole SRR and generates a local *z*-direction magnetic dipole moment, which is also confirmed by the *z* component of the magnetic field distribution. As a result, peak $P_2$ represents a magnetic dipole mode. One of the appealing findings is that the magnetic dipole mode, occurring at about 160 cm$^{-1}$, has a $\lambda/d_2$ ($\lambda$ is the free-space wavelength) ratio as high as ~160. Such high confinement originates from the relatively low carrier density of graphene [11]. As the light frequency increases further, the faster change of oscillation phase divides the original current into three segments and forms a resonance of $P_3$. The surface current density distribution of peak $P_3$ shows the excitation of a quadrupole. Note that both spectra in Fig. 3(a) rise in the high frequency end, which is attributed to the plasmon oscillation cross the ribbon (ring itself is a curved ribbon).

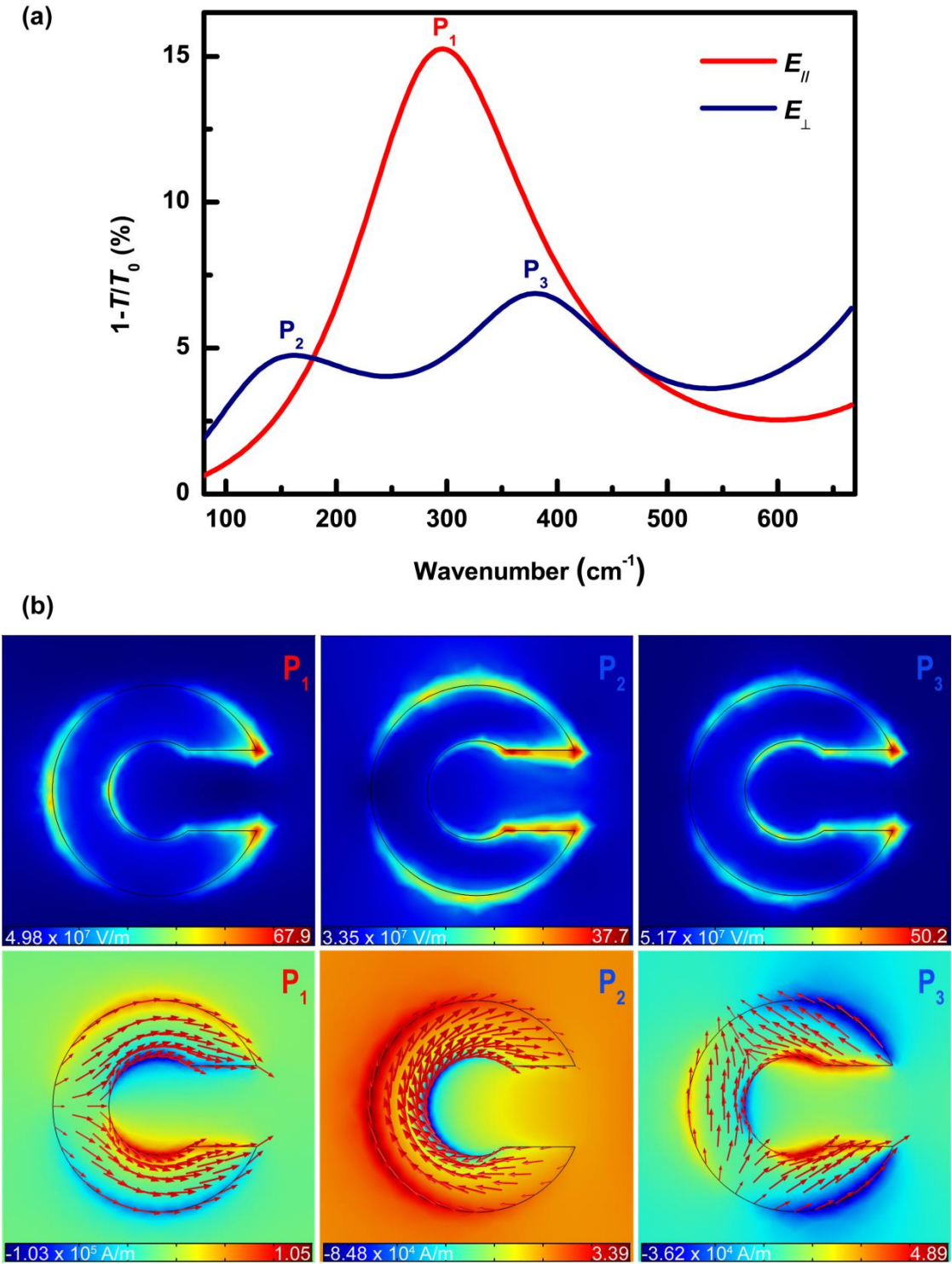

**FIG. 3. Spectra and field simulations.** (a) The simulated extinction spectra of a typical SRR array with two different incident polarization directions. The measured spectra are shown in Fig. 1(d). The resonance modes are denoted as $P_1$, $P_2$ and $P_3$. (b) Simulated surface electric-field magnitude distribution on the upper panels, and $z$-direction magnetic field magnitude ($H_z$) distribution (pseudo-color map) as well as surface current density (arrows) on the lower panels. The values in the scale bars shown in the bottom of the pseudo-color

## C. Tunability of graphene SRRs

One of the attractive advantages of graphene SRR is the high tunability. To investigate the carrier density-dependent spectral response of the SRRs, a chemically doped double-layer graphene SRR array on Si substrate with various doping concentrations was characterized. The SRR array has the same geometrical parameters as that of Fig. 1(d). As shown in Fig. S2, the fitted Drude response before nanofabrication gives the initial carrier density of $9.5 \times 10^{12} \, cm^{-2}$. According to the experimental data shown in Fig. S3, the carrier density after doping is about 3.5 times of the original [27]. Procedures for adjusting the doping level are described in the Experimental section. As shown in Fig. 4(a), with an increasing doping level, all the resonance peaks blueshift, accompanied by the enhanced oscillator strength as expected from the carrier density scaling law in graphene [14]. Fig. 4(b) shows the frequencies of the magnetic dipole $P_2$ and the quadrupole $P_3$ as functions of the frequency of the electric dipole $P_1$. Linear fits work well for the data points, which indicates that three resonance modes follow the same carrier density scaling law. Specifically, the shifts of $P_1$, $P_2$ and $P_3$ are about 80 cm$^{-1}$, 30 cm$^{-1}$ and 100 cm$^{-1}$, which correspond to 27%, 20% and 27% of the initial frequencies of the three modes, respectively. This demonstrates nice tunability through doping. Another graphene SRRs sample with 300 nm outer diameter also exhibits good tunability, as shown in Fig. S4 (see Supplemental Material [27]).

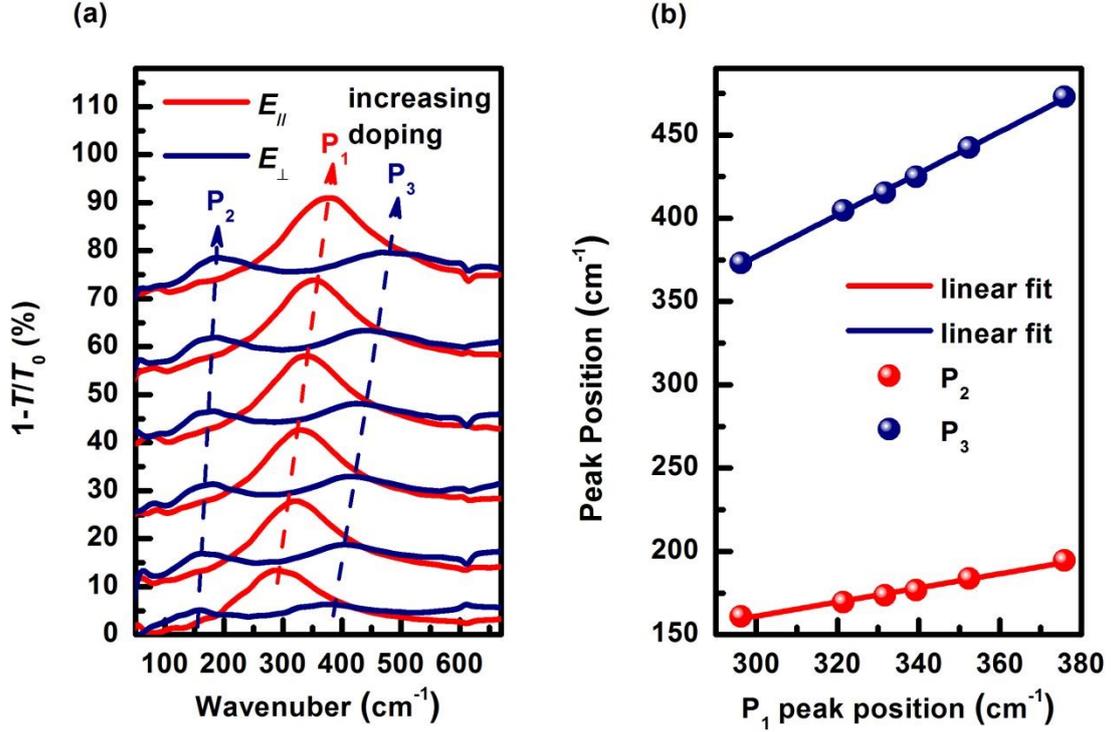

**FIG. 4. Tunability of graphene SRRs on silicon substrate. (a) Extinction spectra of the typical double-layer graphene SRR array with an increasing doping level. This is the same SRR array with $d_2$ of 400 nm in Fig. 2 (a) on Si substrate. Spectra for different doping levels are shifted vertically for clarity. (b) The frequencies of peak $P_2$ and peak $P_3$ (perpendicular polarization) as functions of the frequency of peak $P_1$ (parallel polarization). Solid lines are linear fits.**

In addition to doping, other means can modify the resonances as well. In particular, stacking multiple layers and engineering the plasmon coupling to the substrate phonons are efficient ways [32-34]. We fabricated graphene SRR arrays from one to three graphene layers with the same geometrical parameters ($d_1$ = 100 nm, $d_2$ = 400 nm, $g$ = 100 nm, $p$ = 500 nm) on $SiO_2$/Si substrates, which have surface polar phonons. As shown in the left part of Fig. 5(a), due to the coupling to the substrate polar phonons, those resonance modes split into two branches near the phonon energy [32, 35]. It is worth noting that the dip at ~500 cm$^{-1}$ is regarded as the resonance frequency of a longitudinal phonon mode of $SiO_2$ insulator layer. Meanwhile, with increasing number of stacked graphene layers, the larger effective carrier density leads to higher resonance frequencies [28], as shown in Fig. 5(b). However, the frequency dependence on the layer number follows different trends for different resonance modes, presumably due to the different hybridization effect with the surface phonons. The right part of Fig. 5(a) shows the simulated extinction spectra of the three samples, which excellently agree with the measured ones. The simulated electric field distribution and surface current density distribution of the double-layer graphene SRR array on $SiO_2$/Si substrate, which are used to check the nature of the resonance modes, are shown in Fig. S5 of the Supplemental Material [27]. Besides, another series of extinction spectra of one to three-layer graphene SRR arrays with 450 nm outer diameter are displayed in Fig. S6 of the Supplemental Material [27]. An example verifying the doping tunability of the graphene SRRs on $SiO_2$/Si substrate is shown in Fig. S7 (see Supplemental Material [27]).

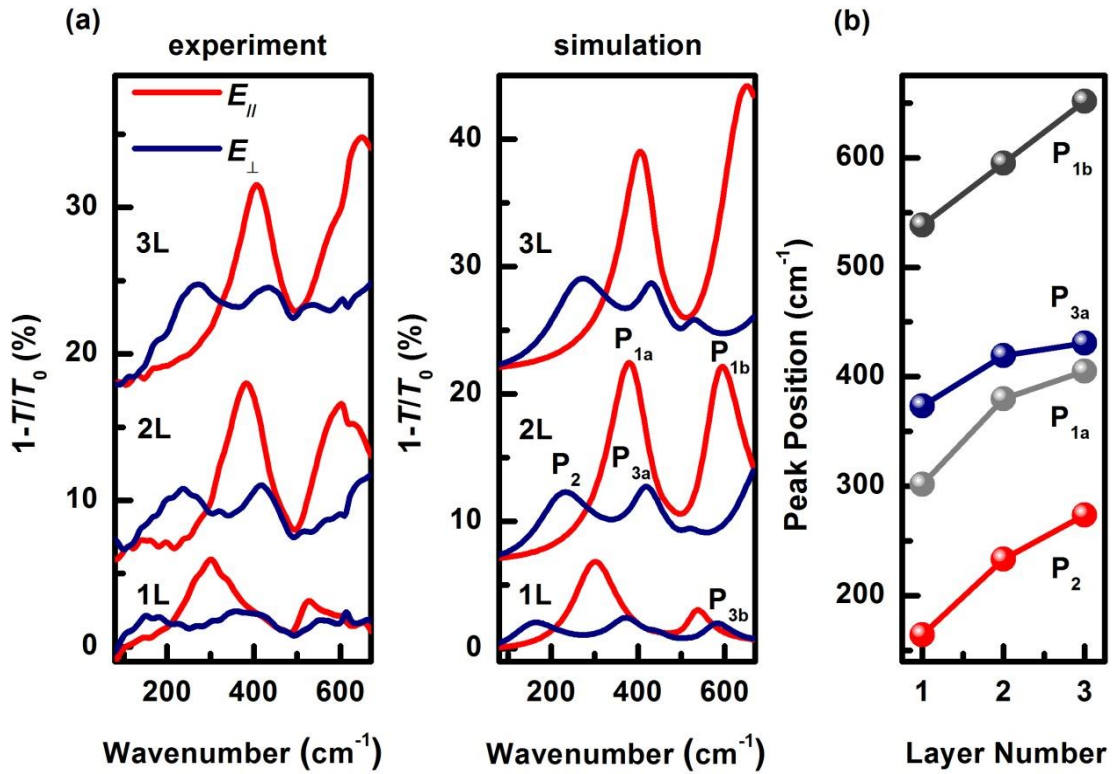

**FIG. 5. SRRs on $SiO_2$/Si substrate with one to three graphene layers. (a) Measured and simulated spectra of one to three layered graphene SRR arrays with the same geometric parameters as that of the SEM image in Fig. 1(b). Spectra are vertically offset for clarity. (b) The frequencies of the peaks (labeled in the simulation spectra) as functions of the layer number of graphene. Note that due to the hybridization with a surface polar phonon mode on $SiO_2$, the original electric dipole $P_1$ and quadrupole $P_3$ split into two branches. Here we**

denote them as $P_{1a}$, $P_{1b}$ and $P_{3a}$, $P_{3b}$ (some of them beyond the scope of our measurement). Since $P_2$ has a relatively low frequency, it shows little hybridization effect with the surface phonon, hence we don't treat it as two branches here.

## IV. CONCLUSION

In conclusion, we have fabricated and characterized graphene SRR arrays with highly confined and widely tunable magnetic dipole, electric dipole and quadrupole modes. In particular, good tunability with chemical doping indicates a potential powerful in-situ tuning with electrical gating, which, of course, requires a more sophisticated device design. Since magnetic response is a fundamental and highly desired property of the SRR, our demonstration of such resonance opens very promising pathways toward tunable and compact magnetic metasurfaces based on graphene in the terahertz and far-infrared regimes.

## ACKNOWLEDGMENTS

H.Y. is grateful to the financial support from National Natural Science Foundation of China (Grant Nos. 11874009, 11734007), the National Key Research and Development Program of China (Grant Nos. 2016YFA0203900 and 2017YFA0303504), Strategic Priority Research Program of Chinese Academy of Sciences (XDB30000000), and the Oriental Scholar Program from Shanghai Municipal Education Commission. C.W. is grateful to the financial support from the National Natural Science Foundation of China (Grant No. 11704075) and China Postdoctoral Science Foundation. L. Z. acknowledges the support from National Natural Science Foundation of China (Grant Nos. 11674068, 11734007), the National Key Research and Development Program of China (Grant No. 2017YFA0303504). X. L. acknowledges the support from National Natural Science Foundation of China (Grant No. 51772043). Part of the experimental work was carried out in Fudan Nanofabrication Lab.